\documentclass[superscriptaddress,twocolumn,showpacs,prl]{revtex4-2}
\usepackage[utf8]{inputenc}
\usepackage{amsmath}
\usepackage{braket}
\usepackage{graphicx}
\usepackage{amsfonts}
\usepackage{pgfplots}
\usepackage{csquotes}
\usepackage{hhline}
\usepackage{amssymb}
\usepackage{listings}
\usepackage{color}
\usepackage{ulem}
\usepackage{booktabs}
\usepackage{comment}
\usepackage{lipsum}
\usepackage{dcolumn}
\usepackage{tabularx}
\usepackage{braket}
\usepackage{float}
\usepackage{caption} 
\usepackage{subcaption}
\makeatletter
\newcommand{\ssymbol}[1]{^{\@fnsymbol{#1}}}
\makeatother
\definecolor{codegreen}{rgb}{0,0.6,0}
\definecolor{codegray}{rgb}{0.5,0.5,0.5}
\definecolor{codepurple}{rgb}{0.58,0,0.82}
\definecolor{backcolour}{rgb}{0.95,0.95,0.92}
\DeclareMathOperator\erf{erf}
\lstdefinestyle{mystyle}{
    backgroundcolor=\color{backcolour},   
    commentstyle=\color{codegreen},
    keywordstyle=\color{magenta},
    numberstyle=\tiny\color{codegray},
    stringstyle=\color{codepurple},
    basicstyle=\footnotesize,
    breakatwhitespace=false,         
    breaklines=true,                 
    captionpos=b,                    
    keepspaces=true,                 
    numbers=left,                    
    numbersep=5pt,                  
    showspaces=false,                
    showstringspaces=false,
    showtabs=false,                  
    tabsize=2
}
 
\usepackage[colorlinks=true,linkcolor=blue,citecolor=blue,urlcolor=blue]{hyperref}

\newcolumntype{C}{>{\centering\arraybackslash}X}
\begin{document}

\author{Aakash Warke}
\email{warkeaakash@gmail.com}
\affiliation{Department of Physics, Johannes Gutenberg Universität Mainz, Institute of Physics, Staudingerweg 7, 55128 Mainz, Germany}

\author{Janis Nötzel}
\affiliation{Emmy Noether Group for Theoretical Quantum Systems Design, Technical University of Munich, D-80333 Munich, Germany}

\author{Kan Takase}
\affiliation{Department of Applied Physics, School of Engineering, The University of Tokyo, 7-3-1 Hongo, Bunkyo-ku, Tokyo 113-8656, Japan}
\affiliation{Optical Quantum Computing Research Team, RIKEN Center for Quantum Computing, 2-1 Hirosawa,
Wako, Saitama 351-0198, Japan}

\author{Warit Asavanant}
\affiliation{Department of Applied Physics, School of Engineering, The University of Tokyo, 7-3-1 Hongo, Bunkyo-ku, Tokyo 113-8656, Japan}
\affiliation{Optical Quantum Computing Research Team, RIKEN Center for Quantum Computing, 2-1 Hirosawa,
Wako, Saitama 351-0198, Japan}

\author{Hironari Nagayoshi}
\affiliation{Department of Applied Physics, School of Engineering, The University of Tokyo, 7-3-1 Hongo, Bunkyo-ku, Tokyo 113-8656, Japan}

\author{Kosuke Fukui}
\affiliation{Department of Applied Physics, School of Engineering, The University of Tokyo, 7-3-1 Hongo, Bunkyo-ku, Tokyo 113-8656, Japan}

\author{Shuntaro Takeda}
\affiliation{Department of Applied Physics, School of Engineering, The University of Tokyo, 7-3-1 Hongo, Bunkyo-ku, Tokyo 113-8656, Japan}

\author{Akira Furusawa}
\email{akiraf@ap.t.u-tokyo.ac.jp}
\affiliation{Department of Applied Physics, School of Engineering, The University of Tokyo, 7-3-1 Hongo, Bunkyo-ku, Tokyo 113-8656, Japan}
\affiliation{Optical Quantum Computing Research Team, RIKEN Center for Quantum Computing, 2-1 Hirosawa,
Wako, Saitama 351-0198, Japan}

\author{Peter van Loock}
\email{loock@uni-mainz.de}
\affiliation{Department of Physics, Johannes Gutenberg Universität Mainz, Institute of Physics, Staudingerweg 7, 55128 Mainz, Germany}


\title{Photonic Quantum Receiver Attaining the Helstrom Bound}

\begin{abstract}
We propose an efficient decomposition scheme for a quantum receiver that attains the Helstrom bound in the low-photon regime for discriminating binary coherent states. Our method, which avoids feedback as used in Dolinar's case, breaks down nonlinear operations into basic gates used in continuous-variable quantum computation. We account for realistic conditions by examining the impact of photon loss and imperfect photon detection, including the presence of dark counts, while presenting squeezing as a technique to mitigate these noise sources and maintain the advantage over SQL. Our scheme motivates testing quantum advantages with cubic-phase gates and designing photonic quantum computers to optimize symbol-by-symbol measurements in optical communication.
\end{abstract}


\maketitle
\captionsetup{justification=raggedright,singlelinecheck=false}

\textit{Introduction.---\label{AJP_Sec1}}
Discriminating binary coherent states with minimal error probability is a fundamental problem in quantum information theory and has significant implications for optical communication systems. The theoretical limit for this discrimination, known as the Helstrom bound, defines the lowest possible error probability achievable by any quantum measurement \cite{CWH1967}. Under coherent state modulation, this Helstrom bound for the binary alphabet $\ket{\pm\alpha}$ with equal \textit{a priori} probabilities, is given by $p:= (1/2)\big(1-\sqrt{1-e^{-4|\alpha|^2}}\big)$. Conventional coherent receivers use homodyne detection and operate at the \textit{Standard Quantum Limit} (SQL), leaving an exponential gap above this Helstrom bound. Although Helstrom did not provide a realization for this lower bound, many approaches approximate this measurement through practical detection schemes \cite{GL2016, MT2008, JG2007, PvL2005, FEB2018}. Kennedy's proposal achieves the same error exponent as the Helstrom bound for sufficiently large $|\alpha|$ \cite{RK1973}, while Dolinar's proposal reaches this limit through adaptive feedback control and time-dependent displacements \cite{SJD1973}. For this particular application, feedback is challenging to implement, because it demands ultra-fast, precise adjustments based on real-time measurements, which are constrained by the physical limits of detector bandwidth, and feedback delay, making it difficult to implement. This raises the question - \textit{Is there a practical, systematic, single-shot method to achieve the Helstrom bound?} 

To address this, Sasaki and Hirota proposed an optimal unitary operation before a standard photon detector that enables one to attain the Helstrom bound for $\ket{\pm\alpha}$ \cite{SH1996}. Realizing this measurement in practical systems indeed excludes feedback from the detector, yet presents challenges due to its inherent complexity and the nonlinear operations required. This demands efficient decomposition of this unitary using the universal continuous-variable (CV) gate set that is available and achievable with current technology, given by $\{\hat{F}, e^{it\hat{x}}, e^{it\hat{x}^2}, e^{it\hat{x}^3} \}$. Here, $\hat{x}$ and $\hat{p}$ are the optical analogues of position and momentum operators, and $\hat{F} = \exp\left[i\frac{\pi}{2}(\hat{x}^2 + \hat{p}^2)\right]$ is the Fourier gate, allowing to switch between $\hat{x}$ and $\hat{p}$. While the linear and quadratic Gaussian gates incorporate optical phase-space displacement, rotation, and squeezing operations, this universal set also includes an essential non-Gaussian element, $e^{it\hat{x}^3}$ -- the cubic phase gate.  It is the simplest and canonical single-mode nonlinear gate that can be added to turn the set of multi-mode Gaussian operations into a gate set universal for quantum computation. This gate leaves the position operator unchanged and transforms the momentum operator as $\hat{p}\rightarrow\hat{p} + \frac{3}{2} t\hat{x}^2$. It was first proposed with the aim of introducing a non-Clifford element to complete the logical universal gate set for Gottesman-Kitaev-Preskill (GKP) qubits \cite{DG2001}. Cubic phase gates can be conditionally prepared using photon measurements on Gaussian states or approximations with Fock states \cite{DG2001, PvL2009, MY2013}. Nonlinear squeezing \cite{AF2021} and nonlinear feedforward operations guided by homodyne detection enable precise measurement of quadratures, strengthening their non-Gaussianity \cite{AF2016}. On combining these operations with non-Gaussian ancilla states, single-mode cubic phase gates can be realized efficiently. These gates are thus being considered important for decomposing unitary transformations into a universal set of gates, enabling applications in photonic quantum computing, such as studying the Bose-Hubbard model \cite{TK2018}, generating GKP states \cite{PvL2024}, and more \cite{HM2020, CW2015, GF2021}.

In this work, we focus on achieving the Helstrom bound by decomposing the first term of optimal unitary derived by Sasaki and Hirota into a series of gates from the universal CV gate set. This decomposition aims to facilitate the implementation of the optimal, minimum probability of error measurement on photonic chips, thereby optimizing the achievable capacities in optical communication systems. We will detail the mathematical framework for this decomposition and discuss the implications for designing and implementing a photonic receiver that attains the Helstrom bound. We follow the convention $\hat{a} = \hat{x} + i\hat{p}$, with $[\hat{x},\hat{p}] = i/2$, where $\hat{a}$ is the annihilation operator.

\textit{Theory.---\label{AJP_Sec2}}
We first evaluate the Hamiltonian derived by Sasaki and Hirota to attain the Helstrom limit for discriminating binary coherent states $\ket{\pm\alpha}$. It acts on these states after first applying a quantum optical displacement gate, $\hat{D}(-\alpha) = e^{\alpha^*\hat{a}-\alpha \hat{a}^\dagger}$, and reads as \cite{SH1996}:
\begin{eqnarray}
\label{Hamil_Eq1}
    \hat{H} = \sum_{l=0}^{M}\frac{(-\hat{a}^\dagger)^l\hat{a}^l}{l!}\sum_{n=1}^{M}d_n\frac{\hat{a}^n}{\sqrt{n!}} - h.c.
\end{eqnarray}
Here, $d_n = (c_n/\sqrt{1-c^{2}_0)}$ with 
\begin{eqnarray}
    c_{n} = \frac{(-2|\alpha|)^n}{\sqrt{n!}} \frac{e^{-2|\alpha|^2}}{\sqrt{\sum_{m=0}^M |\braket{m|-2\alpha}|^2}}.
\end{eqnarray}
For small photon numbers, i.e. $|\alpha| \ll 1$, implying $M = 1$, the Hamiltonian can be expressed as 
$\hat{H} = d_1 \big(\hat a - \hat a^\dagger - \hat a^\dagger \hat a^2 + \hat a^{\dagger 2}\hat a \big)  = 2 d_1\ \big(i\hat{p} - i\hat{x}\hat{p}\hat{x}+\hat{x}\hat{p}^2-\hat{p}^2\hat{x}-i\hat{p}^3\big)$.
These expressions can be further simplified using $[\hat x, \hat p^2] = i \hat p$ and $2 \hat x \hat p \hat x + 2\hat p^3 = \big(2 \hat x^2 \hat p -i \hat x + 2\hat p^3 \big)/2 + \big(2 \hat p \hat x^2 +i \hat x + 2\hat p^3 \big)/2 = \hat x^2 \hat p + \hat p \hat x^2 + 2 \hat p^3$.
The corresponding unitary can then be written as:
\begin{align}
\hat{U}_{NL}(t) &= \exp[(-4i\hat{p} + 2i\hat{p}^3 + i \hat{x}^2\hat{p} + i\hat{p}\hat{x}^2)t ],
\label{Unit_M2}
\end{align}
where $t$ is derived in \cite{SH1996}, demonstrating the trade-off between the optimal interaction time of $\hat{U}_{NL}(t)$ for a fixed mean photon number per mode, and is given by
\begin{eqnarray}
    t=-\arctan\Bigg[\frac{\sqrt{1-e^{-4|\alpha|^2}}-1+e^{-4|\alpha|^2}}{\sqrt{1-e^{-4|\alpha|^2}}+1-e^{-4|\alpha|^2}}\Bigg]^{\frac{1}{2}}.
\end{eqnarray}
The above unitary operation can trivially be split into individual factors:
\begin{align}
\label{UNL}
    \hat{U}_{NL}(t) &= \exp\left[-4it\hat{p}\right]\exp\left[2it\hat{p}^3\right]\exp[it(\hat{x}^2\hat{p} + \hat{p}\hat{x}^2)]\\\nonumber &\quad + \mathcal{O}(t^2).
\end{align}
This method of splitting is certainly not the most efficient, but it is the simplest one. The first two factors in the above equation are already in the form of two CV gates, a simple position displacement, and a cubic gate. The remaining task is to decompose the last factor $\exp[it(\hat{x}^2\hat{p} + \hat{p}\hat{x}^2)]$ into elementary gates known from CV quantum computation which can be potentially implemented, for instance, by employing suitable photonic ancilla states and homodyne measurements with feedforward. One can use commutator approximations to decompose this last factor \cite{SB1999}, but we show its inefficiency in the Supplementary Material for this decomposition, as it accumulates additional error and hence demands more resources. Fortunately, there is an alternative way of dealing with $\exp[it(\hat{x}^2\hat{p} + \hat{p}\hat{x}^2)]$, which does not rely upon commutator approximations. It is based on the observation that a short sequence of elementary gates \textit{sometimes} suffices to \textit{exactly} simulate a desired nonlinear operation \cite{PvL2011}, which is not possible to do considering general frameworks for exact and approximate decompositions \cite{TKJA2019, TKJA2021}. Especially, for Hamiltonians that mix $\hat{x}$ and $\hat{p}$, in general, approximations are needed \cite{PvL2024}. Let us consider the following sequence of operations,
\begin{align}
    &\exp\left[it\hat{x}^3\right] \exp\left[is\hat{p}^2\right] \exp\left[-it\hat{x}^3\right] \\
    &\quad = \exp\left[ is e^{it\hat{x}^3} \hat{p}^2 e^{-it\hat{x}^3} \right] 
    = \exp\left[ is \left( \hat{p} - \frac{3}{2}t\hat{x}^2 \right)^2 \right] \nonumber \\
    &\quad = \exp\left[ is \left( \hat{p}^2 + \frac{9}{4}t^2\hat{x}^4 
    - \frac{3}{2}t(\hat{p}\hat{x}^2 + \hat{x}^2\hat{p}) \right) \right]. \nonumber
\end{align}
Applying this sequence twice, with opposite signs for
$t$ and $s$ for the second time, gives the following result:
\begin{align}
    &\exp\left[it\hat{x}^3\right] \exp\left[is\hat{p}^2\right] \exp\left[-it\hat{x}^3\right] \\
    &\times \exp\left[-it\hat{x}^3\right] \exp\left[-is\hat{p}^2\right] \exp\left[it\hat{x}^3\right] \nonumber \\
    &= \exp\left[ is \left( \hat{p}^2 + \frac{9}{4}t^2\hat{x}^4 
    - \frac{3}{2}t(\hat{p}\hat{x}^2 + \hat{x}^2\hat{p}) \right) \right] \nonumber \\
    &\times \exp\left[ -is \left( \hat{p}^2 + \frac{9}{4}t^2\hat{x}^4 
    + \frac{3}{2}t(\hat{p}\hat{x}^2 + \hat{x}^2\hat{p}) \right) \right] \nonumber \\
    &= \exp\left[ -is(3t(\hat{p}\hat{x}^2 + \hat{x}^2\hat{p})) \right] + \mathcal{O}(s^2). \nonumber
\end{align}
In the final line, we employed straightforward splitting, similar to the method used earlier. Here we can freely choose $t$ and set $t=\frac{1}{3}$. By further changing the sign in front of $s$ and renaming it as $t$, we can reformulate the unitary operation $\hat{U}_{NL}(t)$ in Eq. \eqref{UNL} as follows:
\begin{eqnarray}
\label{U_Decomposed}
\hat{U}(t) &=& \exp\left[-4it\hat{p}\right] \exp\left[2it\hat{p}^3\right] \exp\left[\frac{i}{3}\hat{x}^3\right] 
\exp\left[-it\hat{p}^2\right] \nonumber \\
&\times&\exp\left[-\frac{2i}{3}\hat{x}^3\right] \exp\left[it\hat{p}^2\right] \exp\left[\frac{i}{3}\hat{x}^3\right] 
+\mathcal{O}(t^2). 
\end{eqnarray}
This decomposition has the same order of error
that we had only by splitting the original $\hat{U}_{NL}(t)$ in Eq. \eqref{UNL}. No additional error has been accumulated after decomposing $\exp[it(\hat{x}^2\hat{p} + \hat{p}\hat{x}^2)]$. To obtain a better estimate of Eq. \eqref{Unit_M2}, we concatenate the above decomposition sequence sufficiently many times, leading to short interaction times and an increased number of elementary operations, thereby minimizing the overall error introduced by each approximation step. In this context, the short interaction time achieved in the experimental demonstration of a cubic phase gate in \cite{AF2023} becomes particularly relevant, as the repeated iterations within this decomposition also necessitate similarly reduced interaction times, apart from the phase terms.

\begin{figure}
    \centering
    \includegraphics[scale=0.55]{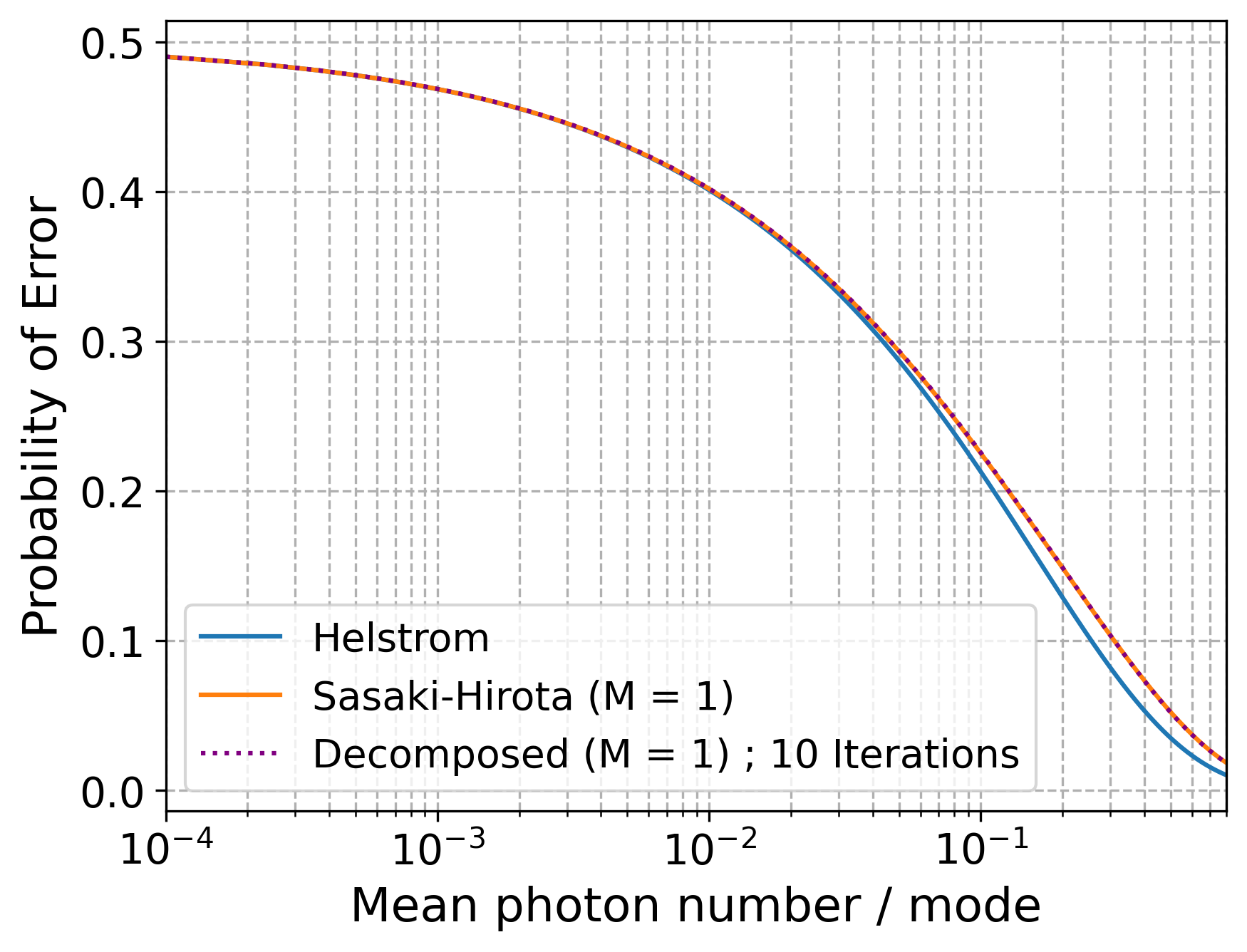}
    \caption{Error probability achieved by our receiver (dotted-purple) over 10 iterations of $\hat{U}(t)$ in Eq. \eqref{U_Decomposed} converges to the Helstrom bound for low mean photon numbers per mode.}
    \label{fig:ProbError_Result}
\end{figure}

A displacement operation $\hat{D}(-\alpha)$, followed by a good approximation of $\hat{U}_{NL}(t)$ obtained via multiple iterations of $\hat{U}(t)$, thus constructs a photonic quantum receiver using the CV gate set capable of distinguishing the binary coherent states $\ket{\pm\alpha}$. The error rate for a given unitary operation $\hat{U}(t)$ before an ideal photon detector with projectors $\{\hat{\Pi}_{\text{off}}, \hat{\Pi}_{\text{on}}\}$ as $\{\ket{0}\bra{0}, \sum_{n=1}^{\infty}\ket{n}\bra{n}\}$ can be derived as \cite{SH1996}:
\begin{align}
\label{ProbErrorExpression}
    P_{error} &= \frac{1}{2} \langle 0 | \hat{U}^{\dagger}(t){\hat{\Pi}_\text{on}}\hat{U}(t)|0 \rangle\\ &+ \frac{1}{2} \langle -2\alpha | \hat{U}^{\dagger}(t)\hat{\Pi}_{\text{off}}\hat{U}(t)| -2\alpha \rangle\nonumber.
\end{align}
Using Eq. \eqref{ProbErrorExpression}, in Fig. \ref{fig:ProbError_Result}, we present a plot comparing the error rates for the $M = 1$ term of the Sasaki-Hirota Hamiltonian, represented by the unitary $\hat{U}_{NL}(t)$, with the error rates of our decomposed photonic scheme. This scheme involves 10 iterations of $\hat{U}(t)$ — comprising 10 linear gates, 40 cubic phase gates, and 20 quadratic gates. The two error rates align optimally while achieving the Helstrom bound in the low-photon regime. Increasing the number of iterations improves the approximation of $\hat{U}_{NL}(t)$, albeit with greater resource demands. One can further expect the induced binary asymmetric channel to reach optimal capacity allowed by symbol-by-symbol measurements.

\begin{figure}
    \centering
    \includegraphics[scale=0.535]{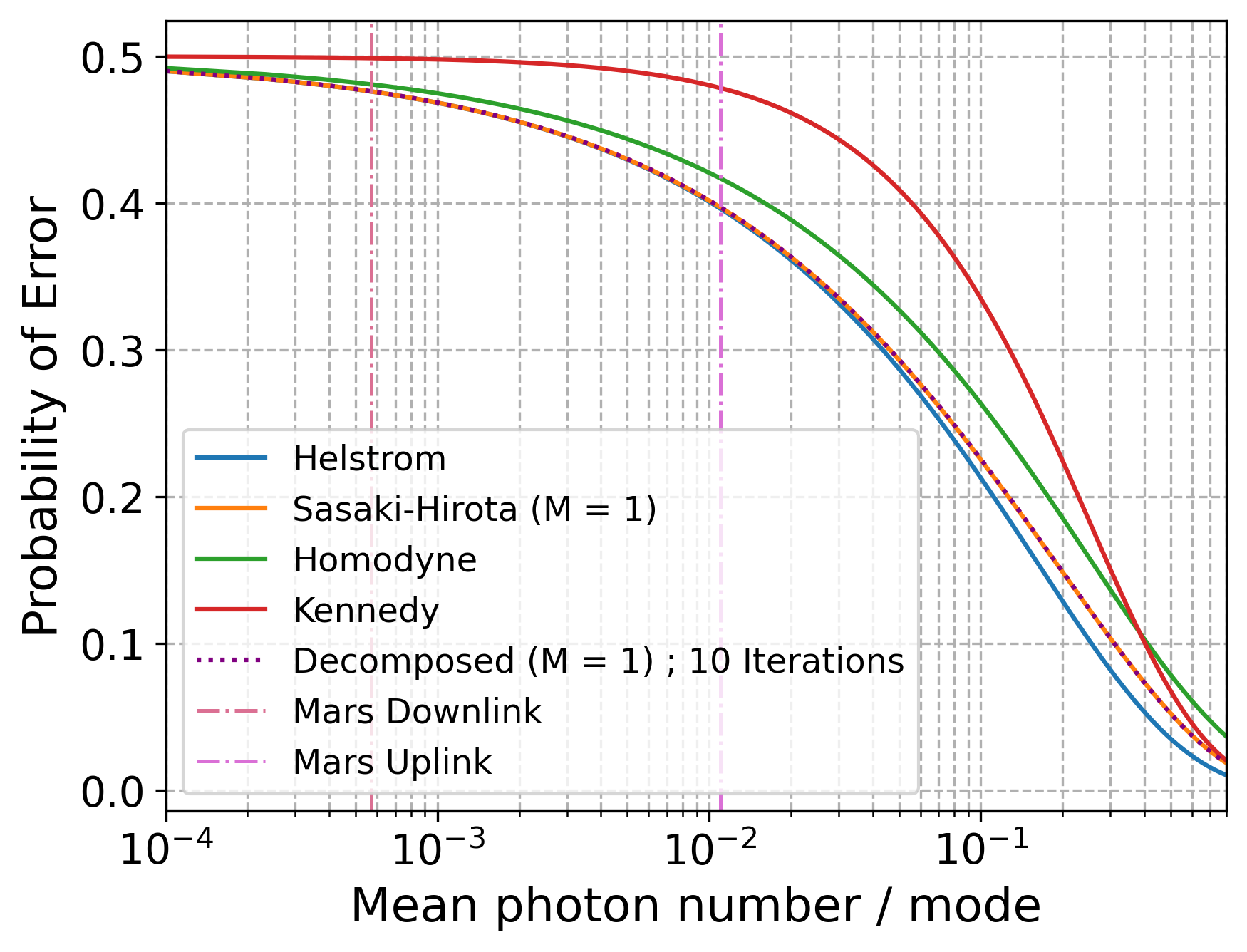}
    \caption{Consistent advantage is observed in error rates of our receiver against Homodyne and Kennedy schemes \cite{RK1973}. We also show the typical values for mean photon numbers for Mars Down/Up-link \cite{SG2022}.}
    \label{fig:Comparison_HomKen}
\end{figure}

\textit{Comparison with Existing Schemes.---\label{AJP_Sec3}}
The simplest receiver commonly implemented in optical communication is the near-optimal homodyne receiver, which outperforms the Kennedy receiver for $|\alpha|^2 \lesssim 0.4$. For the two coherent states $\ket{\pm \alpha}$, the error rate for homodyne detection can be understood through the overlap between the probability distributions of the quadrature measurements and is given by $P_{hom} = (1/2)(1-\erf(\sqrt{2}|\alpha|))$ for equal priors, where $\erf(x)$ is the error function. On the other hand, the Kennedy receiver aims to displace $\ket{\pm\alpha}$ using $\hat{D}(-\alpha)$ and then employs a standard photon detector to decode the outcome. Assuming equal priors, unit efficiency, and no dark counts of this detector, the error rate for the Kennedy receiver is $P_{Ken}=\frac{1}{2}e^{-4|\alpha|^2}$ \cite{RK1973}. We first compare in Fig. \ref{fig:Comparison_HomKen} the performance of our receiver with the homodyne and Kennedy receivers to find a consistent advantage of our scheme. 

\begin{figure}
    \centering
    \includegraphics[scale=0.535]{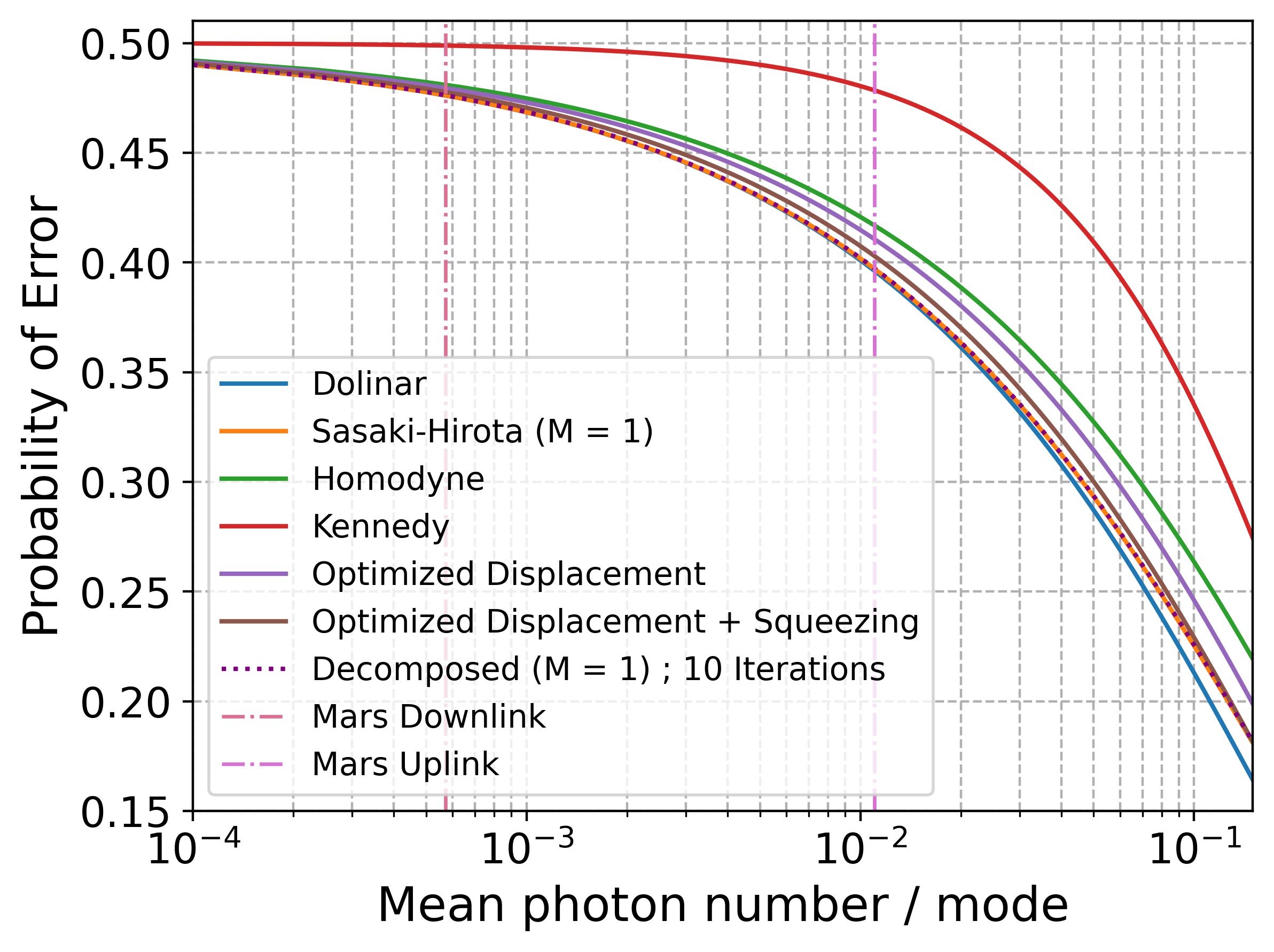}
    \caption{For mean photon numbers greater than 0.15, optimized displacement and squeezing receiver \cite{MT2008} outperforms the error rate of Sasaki-Hirota (M = 1) unitary, and consequently our decomposed scheme.}
    \label{fig:Comparison_All}
\end{figure}

Takeoka and Sasaki proposed using an optimized displacement and a combination of optimized displacement with squeezing as unitary operations before standard photon detectors to construct simple non-Gaussian quantum receivers \cite{MT2008}. We re-derive these results for the ideal case and compare them with our scheme in Fig. \ref{fig:Comparison_All}. We observe a small advantage over the optimized displacement and squeezing receiver for sufficiently low photon numbers and find that this advantage is lost after 0.15 photons/mode. We show in the Supplementary Material that for higher orders of the Sasaki-Hirota Hamiltonian in Eq. \eqref{Hamil_Eq1}, i.e. $M \geq 2$, the corresponding unitary operations result in an error rate lower than all previous receiver designs that employ unitary operations before a photon detector and exclude feedback. This strongly motivates higher-order decompositions of the Hamiltonian for the development of more efficient quantum receivers. 

\begin{figure}
    \centering
    \includegraphics[width=0.98\linewidth]{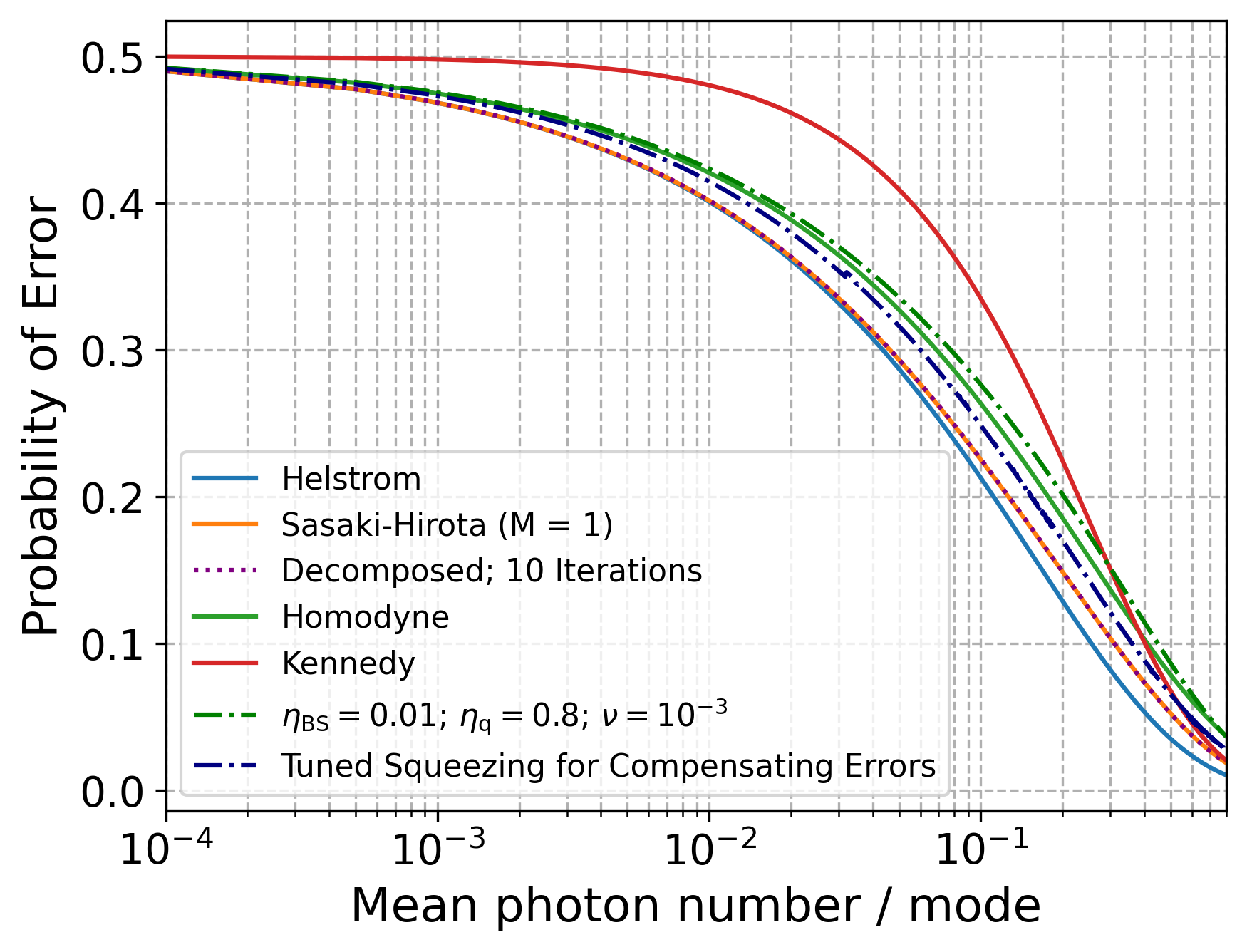}
    \caption{Suitably tuned squeezing shows a reduction in the impact of imperfections, regaining our receiver's advantage against ideal homodyne and Kennedy.}
    \label{fig:TunedSqueezers_CompError}
\end{figure}

\textit{Receiver Imperfections.---\label{AJP_Sec4}}
The optical implementation of a cubic phase gate \cite{AF2016, AF2023} needed for our decompositions is possible with a non-Gaussian ancilla and nonlinear feedforward. Crucially, the nonlinearity is generated solely by a classical nonlinear adaptive control for which unit-fidelity is attainable. The only imperfection will be caused by the ancilla state. Since the receiver is employed in classical optical communication, there is no need for full fault tolerance based on quantum error correction. Nonetheless, we shall model the imperfections related with the ancilla states and the optical detectors by incorporating photon loss into the cubic phase gates as well as the photon detectors. For the latter we include dark counts and a finite quantum efficiency. For non-ideal photon detection, the projector $\hat{\Pi}'_{\text{off}}$ is represented as $e^{-\nu}\sum_{m=0}^{\infty}(1-\eta_q)^m \ket{m}\bra{m}$ and $\hat{\Pi}'_{\text{on}}$ is just $\hat{I} - \hat{\Pi}'_{\text{off}}$ \cite{GJM2004}. The parameters $\nu$ and $\eta_q$ characterize dark counts and quantum efficiency, respectively. We first rewrite Eq. \eqref{ProbErrorExpression} in terms of projectors representing imperfect single-photon detectors $\{\hat{\Pi}'_{\text{off}}, {\hat{\Pi}'_\text{on}}\}$, and then model photon loss through cubic gates. Thus, we couple the output quantum state from the gate with an ancillary vacuum state via a beamsplitter of transmitted amplitude \(\sqrt{1-\eta_{BS}}\) where \(\eta_{BS}\) is the reflectivity of the beamsplitter placed after each cubic phase gate in the sequence $\hat{U}(t)$ for all iterations.
Our model considers each beamsplitter's reflectivity as $\eta_{BS} = 10^{-2}$. 
The linear and quadratic gates are assumed to be ideal with no photon loss. To model imperfect photon detectors, we assume a quantum efficiency of $\eta_q = 0.8$ and a dark count rate of $\nu = 10^{-3}$, aligning with the capabilities of modern superconducting nanowire single-photon detectors (SNSPDs). Using this model for lossy cubic gates and imperfect photon detection, we plot the achievable error rates for different photon numbers (Fig. \ref{fig:TunedSqueezers_CompError}). One effective approach to mitigate these losses is by squeezing the quadrature most affected by photon loss before the cubic phase gate. This improves the signal-to-noise ratio, reducing the impact of photon loss by amplifying the quantum state. We show in Fig. \ref{fig:TunedSqueezers_CompError}, by placing a squeezing operation before every cubic phase gate in $\hat{U}(t)$ in Eq. \eqref{U_Decomposed} and optimizing over the squeezing parameters, that to some extent one can compensate for the aforementioned imperfections. 

\textit{Summary and Discussion.---\label{AJP_Sec5}}
We derived an efficient decomposition of a photonic quantum receiver into CV gates that is, in principle, superior to any known feedback-free receiver scheme in a regime of low mean photon number per mode (less than $\approx 0.15$ photons). In this regime, and for a photon number greater than 0.02, compared with the general and universal commutator-based schemes, the advantage of our receiver-adapted and solely splitting-based decomposition is significant, requiring only tens of gates of reasonable interaction strength. For the experimentally imperfect gates, while demonstrating some form or error mitigation using optical squeezers, we argue that even the hardest elements, the cubic phase gates, will be implementable with unit-fidelity, nonlinear classical feedforward \cite{AF2016, AF2023} and, in principle, near-deterministically when employing recently available photon-number-resolving detectors \cite{DG2001}. While we evaluate decomposition methods for decoding single-mode binary coherent states to attain the Helstrom bound, the scope of designing receivers extends to controlling interactions between temporally collected modes of coherent state inputs, allowing for the extraction of more information from the system than is possible on a per-mode basis through symbol-by-symbol measurement \cite{SG2011, SG2021}. Attaining this advantage involves designing quantum circuits and outer decoders that facilitate multi-modal interactions \cite{SL2011, MW2012}. Realizing superadditivity and attaining the Holevo bound is known to require a non-Gaussian component \cite{SG2014}, which can range from simple photon detectors to more complex multi-mode non-Gaussian gates \cite{HM2024, PvL2024, Furusawa2024}. Additional useful elements may include non-Gaussian photonic circuits followed by homodyne detection \cite{HK2024} and approaches to mixed analog-digital \cite{Liu2024, PvL2024} quantum signal processing. We expect our work to motivate further developments in the design of quantum photonic circuits for optical communication and also, quantum key distribution \cite{ZZ2020, Notarnicola2023}. 


We acknowledge funding from the BMBF in Germany (QR.X, PhotonQ, QuKuK, QuaPhySI), the EU’s Horizon Research and Innovation Actions (CLUSTEC), the DFG Emmy-Noether program (NO 1129/2-1), and the Excellence Track program coordinated by the Mainz Physics Academy. This work was also partly supported by Japan Science and Technology Agency (Moonshot Research and Development)
Grants No. JPMJMS2064 and No. JPMJMS2061, UTokyo Foundation, and donations from Nichia Corporation.

\bibliography{bib}
\newpage
\begin{widetext}
\appendix
\renewcommand{\thesection}{}
\section*{Supplementary Material}
\renewcommand{\thesection}{A} 
\textit{Inefficiency of the Commutator Approximation.---}\label{AppendixA}
Consider the unitary given in Eq. \eqref{UNL}. To handle the last factor $\exp[it(\hat{x}^2\hat{p} + \hat{p}\hat{x}^2)]$, let us now use the usual, general commutator approximation. We will first rewrite $\hat{x}^2\hat{p} + \hat{p}\hat{x}^2$ as $(-4i/6)[\hat{x}^3, \hat{p}^2]$ \cite{PvL2011}. A good feature here is that we are already done with a single commutator, so nested commutators are unnecessary. Approximating nested commutators is notoriously less efficient than approximating commutators. For a commutator approximation, similar to the above splitting, we may also choose the simplest and certainly least efficient way as expressed by:
\begin{eqnarray}
\exp\left[it^2 \left(-\frac{4i}{6}\right) [\hat{x}^3, \hat{p}^2]\right] &=& \exp\left[t^2 \left(\frac{2}{3}\right) [\hat{x}^3, \hat{p}^2]\right] \\
&=& \exp\left[it \left(\frac{2}{3}\right) \hat{p}^2\right] \exp\left[it \hat{x}^3\right] \exp\left[-it \left(\frac{2}{3}\right) \hat{p}^2\right] \exp\left[-it \hat{x}^3\right] + \mathcal{O}(t^3).\nonumber
\end{eqnarray}
Inserting this into the original unitary $\hat{U}_{NL}(t)$ in Eq. \eqref{UNL} gives (replacing $t$ by $t^2$):
\begin{align}
\label{ComApp_UNL}
    \hat{U} &= \exp\left[-4it^2\hat{p}\right] \exp\left[2it^2\hat{p}^3\right] \left\{ \exp\left[it\left(\frac{2}{3}\right)\hat{p}^2\right] \exp\left[it\hat{x}^3\right] \exp\left[-it\left(\frac{2}{3}\right)\hat{p}^2\right]\exp\left[-it\hat{x}^3\right] + \mathcal{O}(t^3)\right\} \\ &\quad + \mathcal{O}(t^4).\nonumber
\end{align}
The first observation here is that the overall error is dominated by the error of the commutator approximation. For example, for \(t \sim 0.1\) we have a "gate strength" of \(t^2 \sim 0.01\) and an error from the commutator approximation of the order of \(t^3 \sim 0.001\) (whereas the error from the splitting is one order of magnitude smaller, \(t^4 \sim 0.0001\)). Concatenating the above decomposition sequence multiple times, which is a must to obtain a gate strength of order unity, leads to amplifying the dominating error from order \(t^3 \sim 0.001\) to order $100t^3 \sim 0.1$ (again assuming \(t^2 \sim 0.01\) and hence 100 iterations). From a quantum information perspective, this error appears too large and so we would have to start with smaller elementary gate strengths. However, as the application in our case is the communication of classical information, we may employ classical methods of error correction and so an error of 10\% may be tolerable. Table \ref{TableA1} depicts the resources required for different interaction times considering additionally, the commutator approximation. Table \ref{TableA2} shows the resource estimates for our original scheme that uses just the splitting approximation. Considering these estimates, we can conclude that our scheme is significantly more resource-efficient for achieving a gate strength of order unity compared to the commutator approximation. However, it is also crucial to recognize that when these decompositions are employed to construct a receiver, the optimal interaction time will depend on the overlap of the states $\ket{\pm\alpha}$. Therefore, it is important to also evaluate the required resources in terms of photon numbers, as illustrated in Fig. \ref{fig:CV_GateCount}.

To achieve a gate strength of unity for constructing a receiver, we found that the commutator approximation derived in Eq. (\ref{ComApp_UNL}) demands more resources compared to our scheme. The resource requirements for the commutator approximation grow exponentially as the mean photon number per mode approaches $\approx 0.02$. In contrast, our scheme, using only the splitting approximation, achieves the same result with significantly fewer gates.
\begin{table}[h]
\centering
\begin{minipage}[t]{0.45\textwidth}
\centering
\resizebox{\columnwidth}{!}{%
\begin{tabular}{|llll|}
\hline
Commutator Approximation                                                                                            &                          &                            &         \\ \hline
\multicolumn{1}{|l|}{\begin{tabular}[c]{@{}l@{}}Largest interaction strength\\ per elementary gate ($t$)\end{tabular}} & \multicolumn{1}{l|}{0.1} & \multicolumn{1}{l|}{0.01}  & 0.001   \\
\multicolumn{1}{|l|}{Linear Gates}                                                                                  & \multicolumn{1}{l|}{100} & \multicolumn{1}{l|}{$10^4$} & $10^6$ \\
\multicolumn{1}{|l|}{Quadratic Gates}                                                                               & \multicolumn{1}{l|}{200} & \multicolumn{1}{l|}{$2\cdot10^4$} & $2\cdot 10^6$ \\
\multicolumn{1}{|l|}{Cubic Gates}                                                                                   & \multicolumn{1}{l|}{300} & \multicolumn{1}{l|}{$3\cdot 10^4$} & $3 \cdot 10^6$ \\ \hline
\end{tabular}%
}
\caption{Resource estimations with respect to interaction time for the commutator approximation following the scaling $\mathcal{O}(1/t^2)$.}
\label{TableA1}
\end{minipage}%
\hspace{0.05\textwidth} 
\begin{minipage}[t]{0.45\textwidth}
\centering
\resizebox{\columnwidth}{!}{%
\begin{tabular}{|llll|}
\hline
Our Scheme                                                                                                          &                          &                           &       \\ \hline
\multicolumn{1}{|l|}{\begin{tabular}[c]{@{}l@{}}Largest interaction strength\\ per elementary gate ($t$)\end{tabular}} & \multicolumn{1}{l|}{0.1} & \multicolumn{1}{l|}{0.01} & 0.001 \\
\multicolumn{1}{|l|}{Linear Gates}                                                                                  & \multicolumn{1}{l|}{10}  & \multicolumn{1}{l|}{100}  & 1000  \\
\multicolumn{1}{|l|}{Quadratic Gates}                                                                               & \multicolumn{1}{l|}{20}  & \multicolumn{1}{l|}{200}  & 2000  \\
\multicolumn{1}{|l|}{Cubic Gates}                                                                                   & \multicolumn{1}{l|}{40}  & \multicolumn{1}{l|}{400}  & 4000  \\ \hline
\end{tabular}%
}
\caption{Resource estimations with respect to interaction time for our scheme following the scaling $\mathcal{O}(1/t)$.}
\label{TableA2}
\end{minipage}
\end{table}
\begin{figure}[H]
    \centering
    \includegraphics[width=0.56\linewidth]{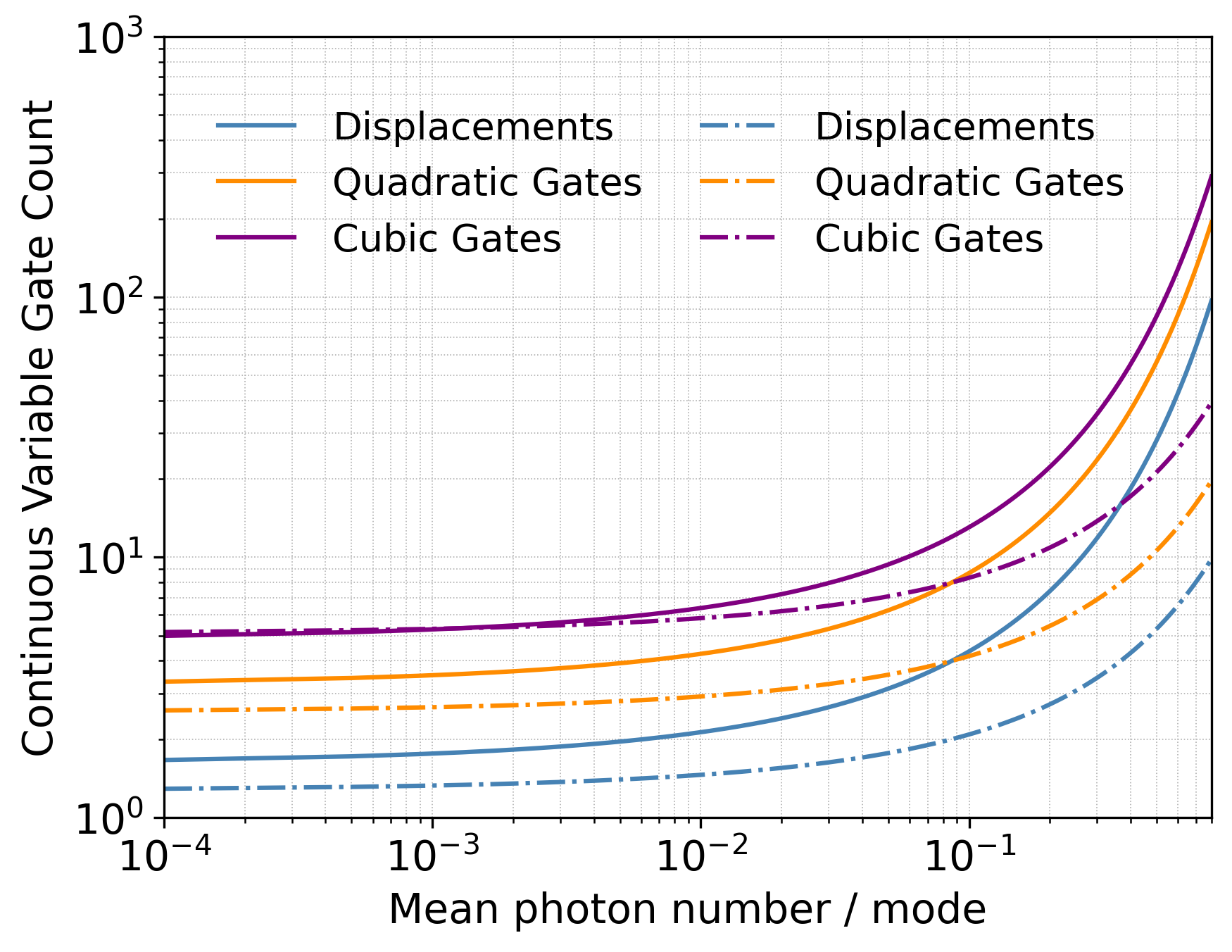}
    \caption{Comparing the CV gate count required for the commutator approximation (thick) to decompose the Sasaki-Hirota Hamiltonian with our scheme (dotted) reveals that our approach uses significantly fewer resources, particularly when exceeding 0.02 photons per mode.}
    \label{fig:CV_GateCount}
\end{figure}

\textit{Higher-order $(M \geq 2)$ Sasaki-Hirota Hamiltonian.---} We show evidence that the inclusion of higher orders of the Sasaki-Hirota Hamiltonian $(M \geq 2)$ results in an error rate lower than previous receiver designs that use unitary operations before a photon detector and exclude feedback. The error rates corresponding to the unitary operations for $M \in \{2, 3, 4, 5\}$ are shown in Fig. \ref{fig:Sasaki_Hirota_Mgeq2}. Notably, $M = 2$ outperforms the optimized displacement and squeezing receiver, suggesting that an efficient decomposition of this and subsequent unitary operations could exceed the performance of previously proposed feedback-free receivers. In the following, we shall look at the Sasaki-Hirota Hamiltonian given in Eq. \eqref{Hamil_Eq1} for $M=2$. First, we write the full Hamiltonian in terms of position and momentum operators as:
\begin{align}
    \hat{H} &= \sum_{l=0}^{M} \frac{(- (\hat{x} - i\hat{p}))^l (\hat{x} + i\hat{p})^l}{l!} \sum_{n=1}^{M} d_n \frac{(\hat{x} + i\hat{p})^n}{\sqrt{n!}} - h.c.\\
    &= \sum_{l=0}^{M} \frac{(- (\hat{x} - i\hat{p}))^l (\hat{x} + i\hat{p})^l}{l!} \sum_{n=1}^{M} d_n \sum_{k=n\%2} \sum_{r=0}^k \binom{k}{r} x^r (ip)^{k-r} \left( \frac{1}{4} \right)^{\frac{n-k}{2}} \frac{n!}{k! \left( \frac{n-k}{2} \right)!} \quad - h.c.
\end{align}
For $M=2$, this expansion takes the form:
\begin{align}
  \hat{H} &= \left(\frac{13}{8} +\frac{\hat{x}^4}{2}+\frac{\hat{p}^4}{2} - 2 \hat{x}^2 - 2\hat{p}^2 +\hat{x}^2\hat{p}^2 -i\hat{x}\hat{p}\right)\cdot \left(d_1\hat{x}+i d_1 \hat{p}+\frac{d_2}{\sqrt{2}}\hat{x}^2-\frac{d_2}{\sqrt{2}}\hat{p}^2+\sqrt{2} d_2 i\hat{x}\hat{p}+\frac{d_2}{2\sqrt{2}}\right) - h.c.\\
  &= d_1 \left(\frac{\hat{x}^5}{2} +\frac{i\hat{p}^5}{2} +\frac{i\hat{x}^4\hat{p}}{2} +\frac{\hat{x}\hat{p}^4}{2} + \frac{13\hat{x}}{8} -\frac{3i\hat{p}}{8} - 2\hat{x}^3 -3i\hat{p}^3 - 4i\hat{x}^2 \hat{p} + 3\hat{x}\hat{p}^2 + \hat{x}^3\hat{p}^2+i\hat{x}^2 \hat{p}^3 -\frac{1}{2}\right)\\
  &\quad + d_2 \Bigg( \frac{13 \hat{x}^2}{8 \sqrt{2}} - \frac{13 \hat{p}^2}{8 \sqrt{2}} + \frac{13 i \sqrt{2} \hat{x} \hat{p}}{8} + \frac{13}{16 \sqrt{2}} + \frac{\hat{x}^6}{2 \sqrt{2}} - \frac{\hat{x}^4 \hat{p}^2}{2 \sqrt{2}} + \frac{i \hat{x}^5 \hat{p}}{\sqrt{2}} + \frac{\hat{x}^4}{4 \sqrt{2}} + \frac{\hat{p}^4 \hat{x}^2}{2 \sqrt{2}} - \frac{\hat{p}^6}{2 \sqrt{2}} + \frac{i \hat{p}^4 \hat{x}\hat{p}}{\sqrt{2}} + \frac{\hat{p}^4}{4 \sqrt{2}} \nonumber\\
  & \quad - \sqrt{2} \hat{x}^4 + \sqrt{2} \hat{x}^2 \hat{p}^2 - 2 \sqrt{2} i \hat{x}^3 \hat{p} - \frac{\hat{x}^2}{\sqrt{2}} - \sqrt{2} \hat{p}^2 \hat{x}^2 + \sqrt{2} \hat{p}^4 - 2 \sqrt{2} i \hat{p}^2 \hat{x} \hat{p} - \frac{\hat{p}^2}{\sqrt{2}} + \frac{\hat{x}^2 \hat{p}^2 \hat{x}^2}{\sqrt{2}} - \frac{\hat{x}^2 \hat{p}^4}{\sqrt{2}} + \sqrt{2} i \hat{x}^2 \hat{p}^2 \hat{x}\hat{p} \nonumber \\
  & \quad + \frac{\hat{x}^2 \hat{p}^2}{2 \sqrt{2}} - \frac{i \hat{x} \hat{p} \hat{x}^2}{\sqrt{2}} + \frac{i \hat{x} \hat{p}^3}{\sqrt{2}} - \sqrt{2} \hat{x} \hat{p}\hat{x} \hat{p} - \frac{i \hat{x} \hat{p}}{2 \sqrt{2}} \nonumber \Bigg) - h.c.
\end{align}
The h.c. term should aid in further simplifying these equations, as the resulting expression is expected to lack terms with only position operators. Although a general scheme for such decompositions could be developed \cite{PvL2024, SB1999, PvL2011, TKJA2019, TKJA2021}, it is deferred to future work. The unitary operations of the above and higher-order Hamiltonians result in unequal transition probabilities from $|\alpha\rangle$ to $|-\alpha\rangle$ and vice-versa, leading to a binary asymmetric channel (BAC). The capacity of this BAC, which depends on the photon number, can be determined analytically and numerically. A common metric for analyzing the capacity's dependence on photon number is the photon information efficiency (PIE), which indicates the number of bits extractable per received photon. Figure \ref{fig:PIE_Result} shows the achievable PIEs for our decomposed scheme and for $M\in \{2,3,4,5\}$. Our analysis here suggests exploring these higher-order unitaries and novel decomposition methods to construct photonic quantum receivers for optical communications is a promising direction.
\begin{figure}
    \centering
    \includegraphics[width=0.57\linewidth]{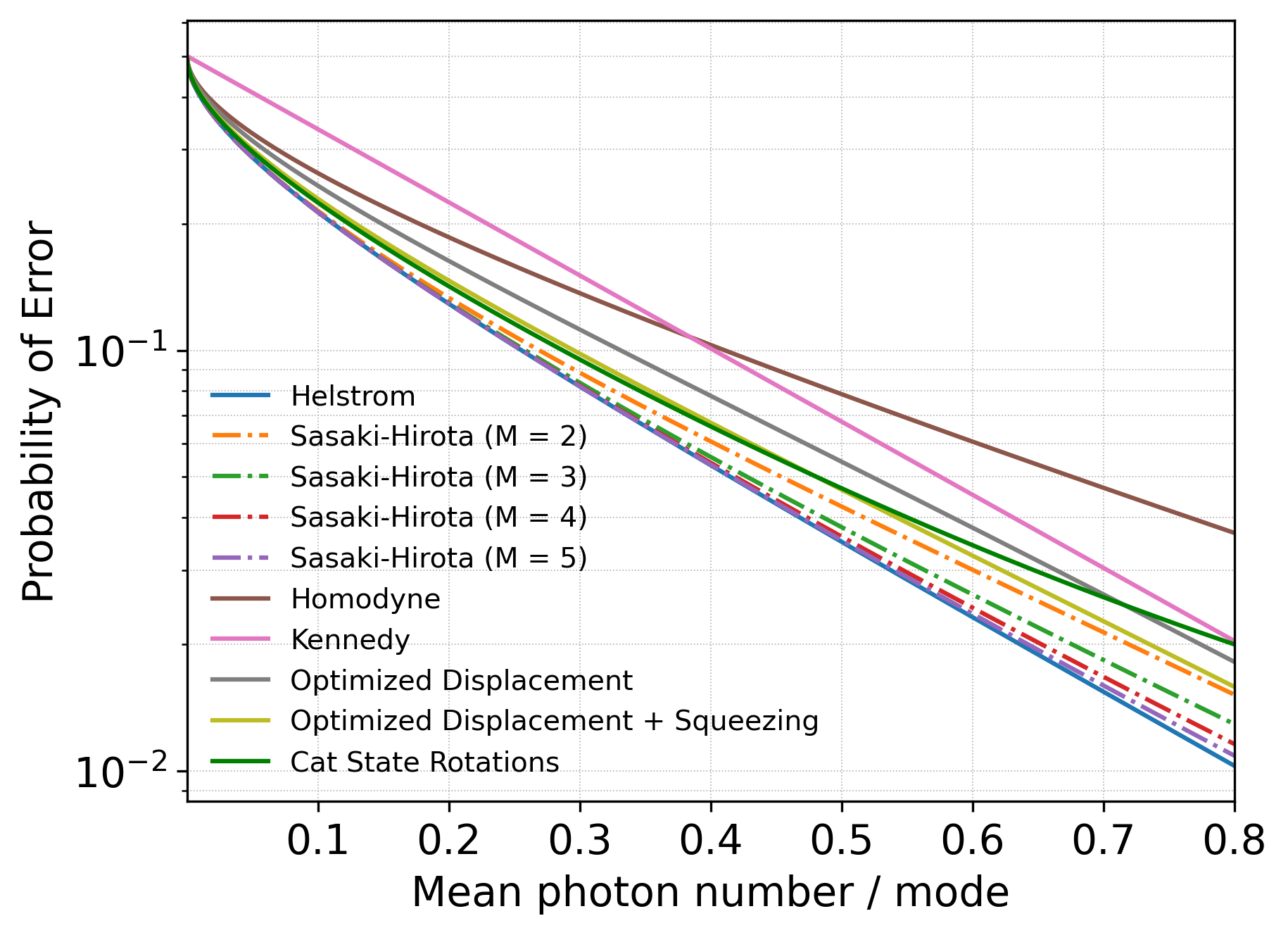}
    \caption{Error rates for higher orders of the Sasaki-Hirota Hamiltonian show improved performance compared to other feedback-free receivers, motivating higher-order decompositions to construct photonic receivers. On comparing our results with the error rates of a recent design using cat state rotations and homodyne detection \cite{HK2024}, we still find photon detection to be a better way of achieving optimality. They also demonstrate the insufficiency of a cubic phase gate in front of the homodyne detector in beating SQL. Cubic phase gates are still useful in this context, as we give an explicit circuit by combining them with Gaussian operations and photon detection to attain the Helstrom bound.}
    \label{fig:Sasaki_Hirota_Mgeq2}
\end{figure}
\begin{figure}
    \centering
    \includegraphics[scale=0.632]{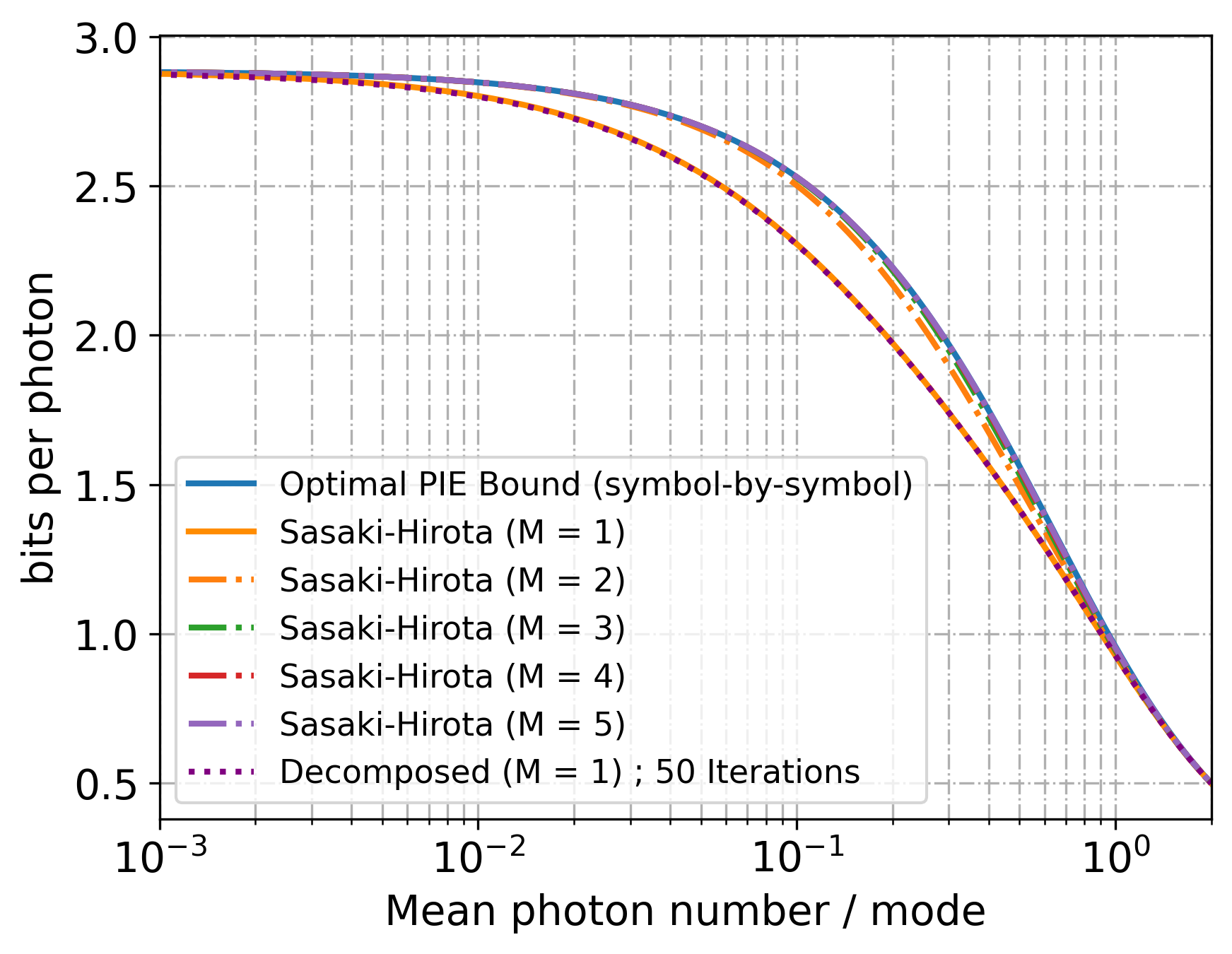}
    \caption{Comparison of achievable photon information efficiencies (PIEs). The optimal PIE bound for symbol-by-symbol measurements is given by \(\frac{1}{\Bar{n}}(1 - H_2(p))\), where \(H_2\) is the binary entropy function and \(\Bar{n}\) is the mean photon number per mode. The plot shows the PIEs for the BAC induced by the Sasaki-Hirota unitary operations of different orders and our receiver for \(M = 1\). Note that Sasaki-Hirota $M \in \{3,4,5\}$ coincide, and so not all the different curves can be resolved in the figure.}
    \label{fig:PIE_Result}
\end{figure} 
\end{widetext}

\end{document}